\documentclass[fleqn,10pt]{wlscirep}
\usepackage[utf8]{inputenc}
\usepackage[T1]{fontenc}
\usepackage{comment}
\usepackage{booktabs}
\usepackage{bookmark} 
\usepackage{subcaption}

\title{Optimization of Connection Patterns between Mobile Phones and Base Stations using Quantum Annealing}

\author[1*]{Taisei Takabayashi}
\author[2]{Shoichi Sudo}
\author[2]{Toshihiro Aoki}
\author[2]{Shun Seo}
\author[1,3,4]{Masayuki Ohzeki}

\affil[1]{Graduate School of Information Sciences, Tohoku University}
\affil[2]{Research Institute of Advanced Technology, SoftBank Corp.}
\affil[3]{Department of Physics, Institute of Science Tokyo}
\affil[4]{Sigma-I Co., Ltd.}

\affil[*]{taisei.takabayashi.s1@dc.tohoku.ac.jp}


\begin{abstract}
\addcontentsline{toc}{section}{Abstract}
In current mobile networks, optimizing which base station a mobile phone in a particular area connects to is crucial for ensuring good communication quality for each mobile phone but presents a challenging combinatorial optimization problem.
In this study, we optimize the connection patterns to base stations using quantum annealing which is a heuristic optimization algorithm using quantum fluctuations.
However, since the number of qubits on a quantum annealer is limited, it is necessary to consider a formulation that efficiently utilizes qubits.
By adopting a variable reduction formulation, we significantly reduce the qubit requirements compared to the naive formulation that is typically used when considering pattern-matching problems. Furthermore, experiments using quantum annealing revealed that the accuracy of the approximate solution obtained by the new formulation is superior to that of the conventional formulation. 
In addition, we demonstrate that the new formulation provides better solutions than the conventional formulation as the problem size increases, even when using simulated annealing, the classical counterpart of quantum annealing.
\end{abstract}
\begin{document} 

\flushbottom
\maketitle

\thispagestyle{empty}

\section*{Introduction}
\addcontentsline{toc}{section}{Introduction}
Each base station adjusts its coverage area in modern mobile networks, but controlling it according to traffic conditions is challenging. When many mobile phones connect to a single base station while there are multiple base stations in the neighborhood, there can be an imbalance in the frequency and power usage among base stations. Therefore, in areas with many mobile phones and multiple base stations, it is necessary to optimize the connection patterns between each phone and base station to ensure communication quality for all mobile phones. 
Various studies have been conducted on optimizing such connection patterns\cite{8007235}.

However, as the number of mobile phones and base stations increases, the number of possible solutions grows exponentially, making the problem difficult to solve. 
Recently, quantum annealing (QA), a method based on the principles of quantum mechanics\cite{Kadowaki1998}, has gained attention as a general approach to solving such combinatorial optimization problems. Quantum annealing searches for the ground state of an Ising model, and since many combinatorial optimization problems can be reduced to finding the ground state of an Ising model\cite{lucas2014ising}, it can be used to solve a wide range of combinatorial optimization problems such as traffic light control\cite{neukart2017traffic, inoue2021traffic, shikanai2023trafficsignaloptimizationusing}, manufacturing\cite{ohzeki2019control, Haba2022}, finance\cite{rosenberg2016solving, venturelli2019reverse}, steel manufacturing\cite{yonaga_quantum_2022}, and algorithms in machine learning\cite{Amin2018, o2018nonnegative, sato_assessment_2021, Urushibata2022, hasegawa2023, goto2023onlinecalibrationschemetraining}. 
With these backgrounds, QA has recently attracted attention.
In this study, we apply QA to optimize connection patterns and demonstrate that QA can be applied to optimization problems in the wireless communication field. 

The quantum annealer, developed by D-Wave Systems, is a representative example of a quantum annealer that implements QA. However, the number of qubits on the D-Wave quantum annealer is still insufficient to solve large-scale real-world problems. 
Moreover, quantum annealers like the D-Wave quantum annealer can only handle optimization problems formulated as quadratic unconstrained binary optimization (QUBO)\cite{glover_quantum_2022}, equivalent to the Ising model. 
Therefore, combinatorial optimization problems must be formulated as QUBO to be solved. 
However, qubits are not fully interconnected on the hardware of the D-Wave quantum annealer, so it is not always possible to map logical variables directly onto qubits. 
If the graph representing the logical variables and their interactions is dense, it may require multiple qubits to represent a single logical variable on the hardware graph, a technique known as minor embedding. 
When using the D-Wave quantum annealer, this embedding process often increases the number of qubits required compared to the number of logical variables. 
Therefore, developing a QUBO formulation that minimizes the required number of qubits is necessary.

A new QUBO formulation that reduces the number of qubits has been proposed in the context of evacuation optimization\cite{ohzeki2023evacuation}. 
In evacuation optimization, the problem is to decide which shelter to assign evacuees to within an evacuation area, given the distance from their current location to each shelter and the capacity limits of the shelters. 
Each evacuee should be assigned to the closest shelter, but due to capacity constraints, not all evacuees can go to their closest shelter. 
However, evacuees can not always take a long way to shelter to rigorously satisfy the capacity constraints.
In addition, the quantum annealer does not have enough capability to optimize the cost function with several strong coefficients, as in the penalty method.
The new formulation thus omits the penalty method and takes binary selection, i.e., the closest or the second one.
As a result, the new formulation proposed for this problem effectively reduces the number of required qubits, allowing larger problems to be addressed on the D-Wave quantum annealer than the naive formulation used for assignment problems.

Recognizing the similarity between the evacuation optimization problem and our problem of optimizing connection patterns to base stations, we employ this new QUBO formulation to address our problem.
We show that this proposed formulation reduces the number of qubits needed for embedding the QUBO and improves the accuracy of the solutions obtained from the D-Wave quantum annealer. 
Furthermore, we confirm that the proposed formulation provides better solutions as the problem size increases, even when using simulated annealing (SA)\cite{Kirkpatrick1983}, the classical counterpart of quantum annealing.
The results suggest that the proposed formulation yields more accurate approximate solutions as the problem size increases, even when using classical QUBO solvers like SA.

The remainder of this paper is organized as follows. In the next chapter, we describe the problem setting for optimizing connection patterns to base stations, including the settings for radio waves emitted from the base stations and the optimization metrics. We introduce both a naive  QUBO formulation for connection pattern optimization and a new QUBO formulation in Chapter 3. In Chapter 4, we present the results of comparing the two formulations regarding the number of qubits required and the accuracy of the solutions obtained on the D-Wave quantum annealer. We also investigate the results of comparing the formulations using SA. Finally, Chapter 5 discusses the experimental results and future research directions.

\section*{Problem Setting}
\begin{figure}[htbp]
  \centering
    \includegraphics[width=98.27mm]{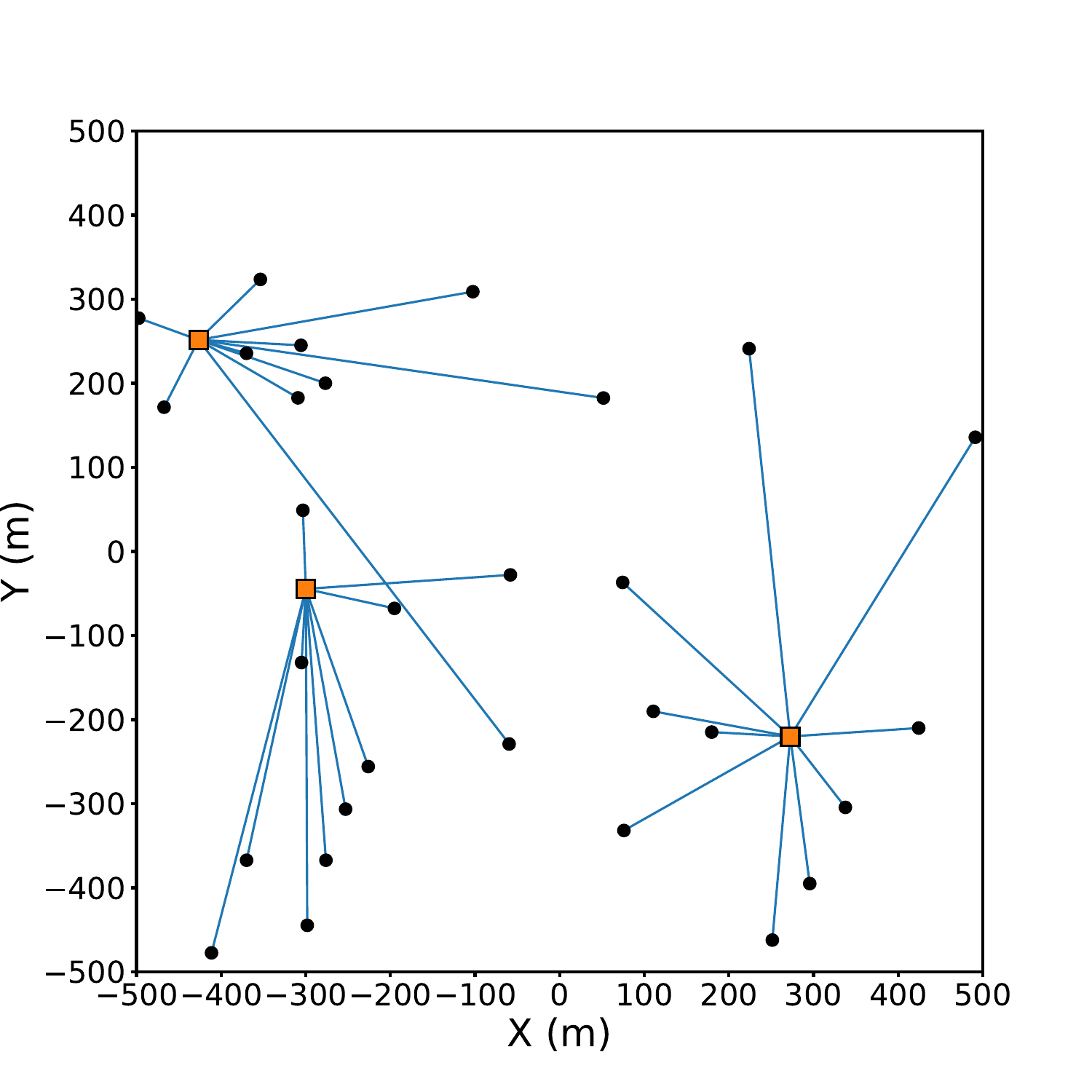}
    \caption{Example of a connection pattern between base stations and mobile phones. Orange squares represent base stations, and black circles show mobile phones. The blue line illustrates the connection between each mobile phone and its corresponding base station.}
    \label{example of assignment}
\end{figure}

\addcontentsline{toc}{section}{Problem Setting}
This chapter describes the problem setting for optimizing connection patterns between mobile phones and base stations. 
As shown in Fig. \ref{example of assignment}, we consider a
situation 
where the positions of mobile phones and base stations are fixed within a certain area. 

There are two constraints for determining the connection pattern. 
The first constraint is that each mobile phone must connect to one of the base stations to enable communication. 
The second constraint is that each base station has a connection limit to prevent overcrowding at certain base stations. 
Under these two constraints, we aim to connect each mobile phone to a base station that maximizes communication quality, which is defined next.

In this study, we use the down-link signal-to-interference-and-noise ratio (DL-SINR) as the metric of communication quality for each mobile phone. 
The following equation expresses DL-SINR:
\begin{equation}
    {\rm DL-SINR} = \frac{S}{I+N}, 
\end{equation}

where $S$ is the desired signal strength, $I$ is the interference signal strength, and $N$ is the noise signal strength. The desired signal strength is the strength of the received signal from the base station to which the mobile phone is connected. The interference signal strength is the sum of the received signal strengths from all other base stations. DL-SINR indicates how a mobile phone can receive data, with higher values representing better communication quality. This paper only considers the down-link, so we refer to DL-SINR simply as SINR.

The following equation calculates the received signal strength:
 \begin{equation}
    {\rm Received \ Signal\  Strength} = \frac{({\rm Transmission}\ {\rm Power}) \times ({\rm Transmission\  Antenna\ Gain})\times({\rm Received\ Antenna\ Gain})}{\rm Path\ Loss}. 
\end{equation}

This study assumes the received antenna gain is a constant value of 1. 
Since the same transmit power is used for all cases, it will be omitted in subsequent calculations. 
Therefore, the received signal strength is calculated based on the path loss and Transmission antenna gain. Path loss is given by the free space path loss (FSPL) as follows:
\begin{equation}
    {\rm FSPL} = \left(\frac{4\pi d f}{c} \right)^2,
\end{equation}
where $\pi$ is the ratio of the circumference of a circle to its diameter, 
$d$ is the distance, $f$ is the frequency, and $c$ is the speed of light. FSPL represents the loss when electromagnetic waves travel through space without obstacles. 
Specifically, it refers to the phenomenon where electromagnetic waves weaken as they travel through space, and it increases as the distance between the mobile phone and the base station increases.

Regarding the Transmission antenna pattern, we consider the following two representative patterns:

\subsection*{Isotropic}
\addcontentsline{toc}{subsection}{Isotropic}
The isotropic pattern is an antenna pattern where beams are emitted uniformly in all directions. Fig. \ref{SINR heat map}(a) is a heat map that shows the SINR at each coordinate when two base stations with isotropic patterns are placed 1 km apart, with mobile phones uniformly distributed every meter. 
All mobile phones at each position connect to the base station that provides the highest SINR. The figure shows that radio waves are emitted uniformly in all directions from the base station in an isotropic pattern, and the SINR decreases with distance from the base station. 
Therefore, in the isotropic pattern, each mobile phone can obtain a high SINR by connecting to the base station closest to it.

\subsection*{Gaussian}
\addcontentsline{toc}{subsection}{Gaussian}
The Gaussian pattern is an antenna pattern with directivity, unlike the isotropic pattern. The gain $G(\phi)$ of the Gaussian pattern is calculated as follows, where $\phi$ is the azimuth angle:
\begin{equation}
    G(\phi) = \max \{G_{\rm main} (\phi), {\rm SLL} \}.
\end{equation}
Here, $G_{\rm main} (\phi)$ denotes the main beam level, while SLL stands for the side lobe level. The main beam level refers to the peak gain of the antenna pattern, which corresponds to the maximum radiation intensity. On the other hand, the sidelobe level represents the radiation intensity in directions other than the main beam, and it is typically weaker than the main beam level. $G_{\rm main}$ can be written as follows, where $G_{\rm max}$ is the maximum gain:
\begin{equation}
    G_{\rm main} (\phi) = \exp \left\{ -\frac{1}{2} \left( \frac{\phi}{\sigma} \right)^2 \times 10^{\frac{G_{\rm max}}{10}} \right\}.
\end{equation}
The half-beam width is defined as follows:
\begin{equation}
    \theta = \frac{\sigma}{2\sqrt{\ln{2}}}.
\end{equation}

\begin{figure}[htbp]
  \begin{minipage}[b]{0.45\linewidth}
    \centering
    \includegraphics[keepaspectratio, scale=0.5]{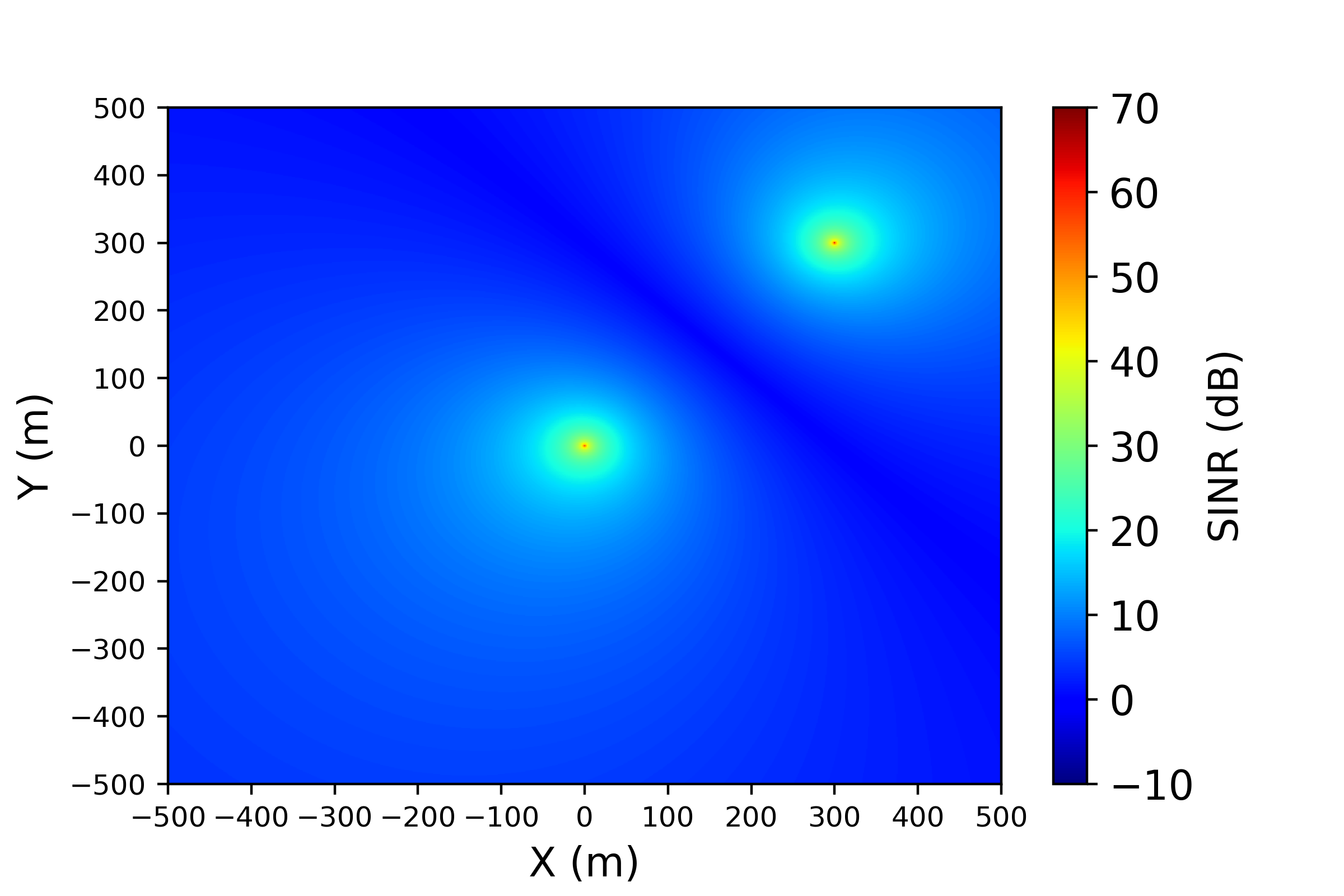}
    \subcaption{isotropic}
  \end{minipage}
  \begin{minipage}[b]{0.45\linewidth}
    \centering
    \includegraphics[keepaspectratio, scale=0.5]{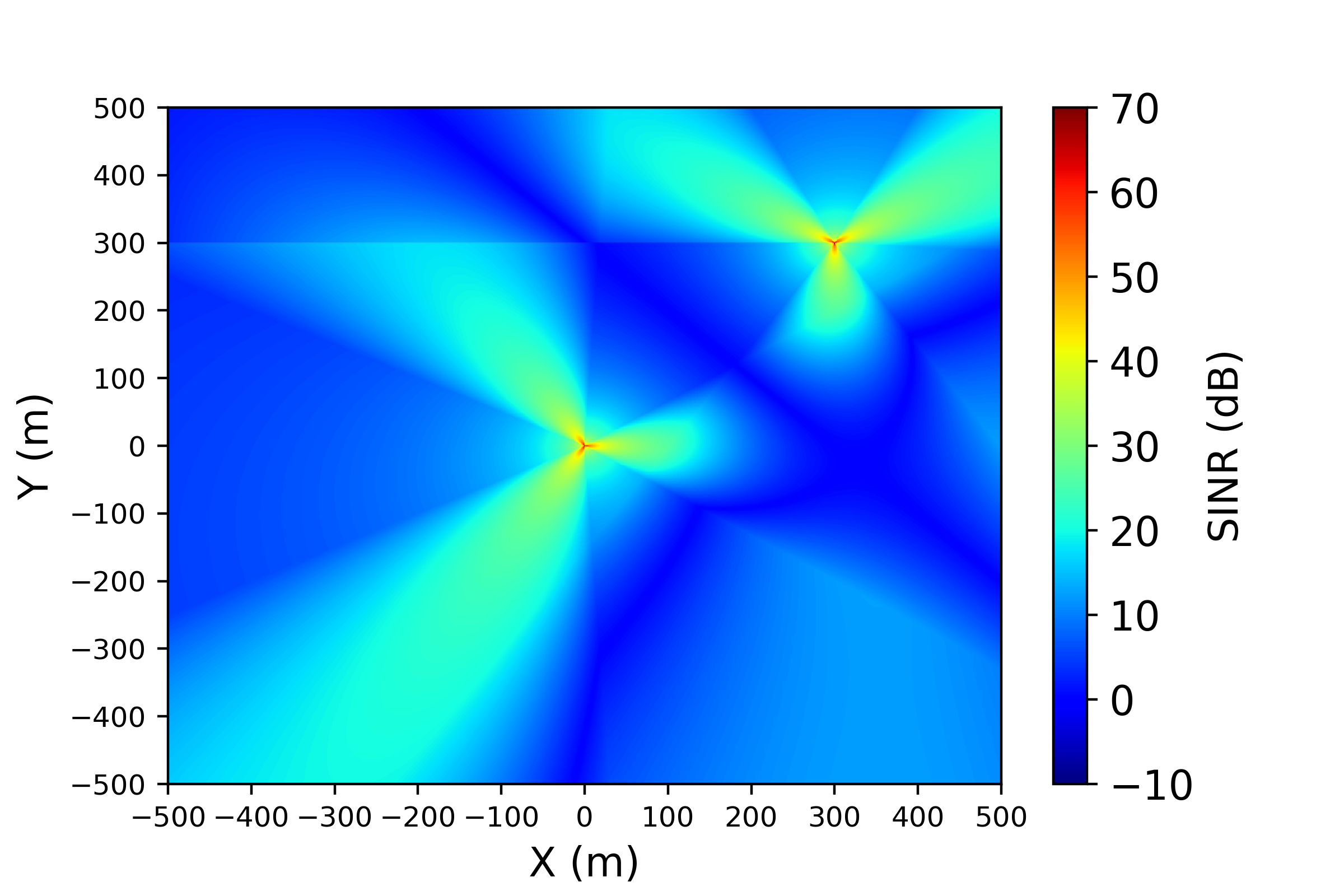}
    \subcaption{Gaussian}
  \end{minipage}
  \caption{Two Transmission antenna patterns. Each heat map shows the SINR at each coordinate when the base station is located at (0, 0) and (350, 350).}
    \label{SINR heat map}
\end{figure}

Fig. \ref{SINR heat map}(b) is the heat map that shows the SINR at each coordinate when two base stations with Gaussian patterns are placed 1 km apart, with mobile phones uniformly distributed every meter. 
All mobile phones at each position connect to the base station that provides the highest SINR. The half-power beam width is 30 degrees, the maximum gain is 0 dB, and the side lobe level is -15 dB. 
The figure shows that in the Gaussian pattern, radio waves are emitted with directivity toward several directions from the base station, and mobile phones can obtain a high SINR by connecting to the base station whose beam is directed toward them, rather than simply connecting to the nearest base station. The beam direction at the actual base station is determined to maximize the SINR in the area, taking into account the beam direction of neighbors. When a new base station is deployed, the existing stations keep its fixed beams while the new one aligns its beams to maximize aggregate SINR.

In connection pattern optimization, we aim to maximize the SINR received by each mobile phone within the area. 
However, since connection limits exist on the base stations, not all mobile phones can connect to the base station that provides the highest SINR. Therefore, this capacity constraint must be included in the formulation. 
The next chapter introduces two QUBO formulations for connection pattern optimization. The first is a naive formulation commonly used for considering pattern-matching. The second applies the QUBO formulation proposed in the previous study\cite{ohzeki2023evacuation}, which reduces the number of qubits required by focusing on the top two best connections for each mobile phone.

\section*{Method}
\addcontentsline{toc}{section}{Method}

\subsection*{Naive formulation}
\addcontentsline{toc}{subsection}{Naive formulation}
In this section, we describe the naive formulation for the problem of maximizing the SINR of each mobile phone when the positions of mobile phones, base stations, and the antenna patterns are fixed. This problem involves determining which mobile phone should connect to which base station. In such cases, the decision variables are generally defined as follows:
\begin{equation}
{x_{i,a}=}
\left\{ \,
    \begin{aligned}
    & 1 \quad {\rm if} \quad {\rm mobile\ phone}\ i \ {\rm connects\ to\ base\ station}\ a. \\
    & 0 \quad {\rm otherwise}.
    \end{aligned}
\right.
\label{naive decision variable}
\end{equation}
By using these variables, we can write the problem that maximizes the sum of the SINR of each mobile phone as follows:
\begin{subequations}
    \begin{equation}
    \max_{x} \quad \sum_{i=1}^{N} \sum_{a=1}^{M} S_{i,a} x_{i,a},
    \end{equation}
    \label{naive objective function}
    \begin{equation}
    {\rm s.t.} \quad \sum_{a=1}^{M} x_{i,a} = 1, \quad \forall i\in \{ 1, \cdots, N\},
    \label{naive constraint 1}
    \end{equation}
    \begin{equation}
    {\rm s.t.} \quad \sum_{i=1}^{N} x_{i,a} = C_a, \quad \forall a\in \{1, \cdots, M\},
    \label{naive constraint 2}
    \end{equation}
\end{subequations}
where $N$ is the number of mobile phones, $M$ is the number of base stations, $S_{i,a}$ is the SINR when mobile phone $i$ connects to base station $a$, and $C_a$ is the capacity of base station $a$. 
Here, $S_{i,a}$ can be uniquely calculated from the fixed positions of the mobile phones and base stations. 
In this paper, we assume $C_a = N / M$ for all $a$, meaning that each of the $N$ mobile phones is connected to one of the $M$ base stations. Eq. \eqref{naive objective function} represents the objective function, which maximizes the total SINR. 
Equation \eqref{naive constraint 1} represents a one-hot constraint, indicating that each mobile phone connects to exactly one base station. 
Equation \eqref{naive constraint 2} denotes the capacity constraint, ensuring that exactly $C_a$ mobile phones are connected to each of the $M$ base stations.

To solve the above problem using a quantum annealer, it needs to be formulated as a QUBO problem. In this formulation, the equality constraints expressed by Eqs. \eqref{naive constraint 1} and \eqref{naive constraint 2} are included in the objective function as penalty terms. 
The resulting QUBO can be expressed as:
\begin{equation}
    \min_{x} \quad -\sum_{i=1}^{N} \sum_{a=1}^{M} S_{i,a} x_{i,a} + \lambda_{1} \sum_{i=1}^{N}  \left( \sum_{a=1}^{M} x_{i,a} - 1\right)^2 + \lambda_{2} \sum_{a=1}^{M}  \left( \sum_{i=1}^{N} x_{i,a} - C_a \right)^2,
\label{naive qubo}
\end{equation}
Here, $\lambda_1$ and $\lambda_2$ are hyper-parameters that control the magnitude of the constraints. 
Since this QUBO formulation considers all possible connection patterns between mobile phones and base stations,
if optimized correctly, it can yield the optimal solution. 
However, solving this naive QUBO using a quantum annealer is not efficient. 
This formulation requires many logical variables, $N\times M$, as shown in Eq. \eqref{naive decision variable}. 
Furthermore, due to the one-hot constraint \eqref{naive constraint 1}, only $N$ of these variables are meaningful, while the remaining $N(M-1)$ variables are redundant. This is not ideal when using a quantum annealer with limited computational resources. To overcome this issue, we introduce a new formulation that effectively reduces the number of logical variables in the next section.

\subsection*{Proposed formulation}
\addcontentsline{toc}{subsection}{Proposed formulation}
We employ a new formulation, originally proposed in the context of evacuation optimization, to effectively reduce the number of logical variables\cite{ohzeki2023evacuation}. 
In evacuation optimization, the problem is to decide which shelter to assign evacuees to within an evacuation area, given the distance from their current location to each shelter and the capacity limits of the shelters. 
Each evacuee should be assigned to the closest shelter, but due to capacity constraints, not all evacuees can go to their closest shelter. This problem setting is analogous to our problem of optimizing connection patterns, where the "magnitude of the SINR when mobile phone $i$ connects to base station $a$" corresponds to the "smallness of the distance between evacuee $i$ and shelter $a$." 

Based on the new formulation proposed in previous research, we define the decision variables as follows:
\begin{equation}
{x_{i}=}
\left\{ \,
    \begin{aligned}
    & 1 \quad {\rm if} \quad {\rm mobile\ phone}\ i \ {\rm connects\ to\ the\ base\ station\ that\ provides\ the\ highest\ SINR}. \\
    & 0 \quad {\rm if} \quad {\rm mobile\ phone}\ i \ {\rm connects\ to\ the\ base\ station\ that\ provides\ the\ second\ highest\ SINR}.
    \end{aligned}
\right.
\label{proposed decision variable}
\end{equation}

In this formulation, each mobile phone only has the option to connect to either the best or the second-best base station, and it is not assigned to any base station that provides lower communication quality. 
This is based on the practical assumption that each mobile phone should ideally be assigned to the best base station, but due to capacity constraints, it may have to settle for the second-best base station.
By formulating the problem this way, the one-hot constraint \eqref{naive constraint 1} of the naive formulation becomes unnecessary. Therefore, we only need to consider the objective function and the capacity constraint, as expressed below:

\begin{subequations}
    \begin{equation}
    \max_{x} \quad \sum_{i=1}^{N}  \left(S^{'}_{i,1} x_{i} + S^{'}_{i,2} (1 - x_{i}) \right),
    \end{equation}
    \label{proposed objective function}
    \begin{equation}
    {\rm s.t.} \quad \sum_{i\in N_1(a)} x_i + \sum_{i\in N_2(a)} (1 - x_i) = C_a, \quad \forall a\in \{1, \cdots, M\},
    \label{proposed constraint 1}
    \end{equation}
\end{subequations}
where $S_{i,1}^{'}$ denotes the SINR achieved when mobile phone $i$ connects to its best base station, whereas $S_{i,2}^{'}$ refers to the SINR obtained when the mobile phone $i$ connects to its second-best base station. The set $N_1 (a)$ comprises all mobiles for which base station $a$ is the best choice, and $N_2 (a)$ comprises those for which base station $a$ is the second-best choice. With these definitions, the problem can be cast as the following QUBO:
\begin{equation}
    \min_{x} \quad - \sum_{i=1}^{N}  \left(S^{'}_{i,1} x_{i} + S^{'}_{i,2} (1 - x_{i}) \right) + \lambda^{'} \sum_{a=1}^{M} \left( \sum_{i\in N_1(a)} x_i + \sum_{i\in N_2(a)} (1 - x_i) - C_a \right)^2. 
\label{proposed qubo}
\end{equation}
In this formulation, as shown in Eq. \eqref{proposed decision variable}, the number of logical variables is reduced to $N$. 
Thus, compared to the naive formulation, the number of logical variables is reduced to $1/M$, significantly decreasing the number of qubits required in the quantum annealer. 
In the naive formulation, many of the variables represent $x_{i,a}=0$, which are redundant, whereas, in the proposed formulation, all $N$ variables directly represent the connection choices of the mobile phones, with no waste in variables. 
Moreover, since each mobile phone is only connected to the best or second-best base station, the proposed formulation is expected to provide a good approximate solution. 
While the naive formulation, which considers all connection patterns, may yield the optimal solution, the proposed formulation is more efficient regarding qubit usage and will likely provide a good approximate solution quickly, especially considering the error-prone nature of QA with an increasing number of qubits.

In the next chapter, we experimentally compare the effectiveness of the proposed formulation with the naive formulation. 
Specifically, we compare the number of qubits required and the accuracy of the solutions obtained from the D-Wave quantum annealer for the same problem size in both formulations. In addition, we used SA instead of D-Wave quantum annealers to investigate the dependence of the accuracy of the solutions obtained by the two formulations on the number of mobile phones.

\section*{Experiments}
\begin{figure}[htbp]
  \begin{minipage}[b]{0.45\linewidth}
    \centering
    \includegraphics[keepaspectratio, scale=0.3]{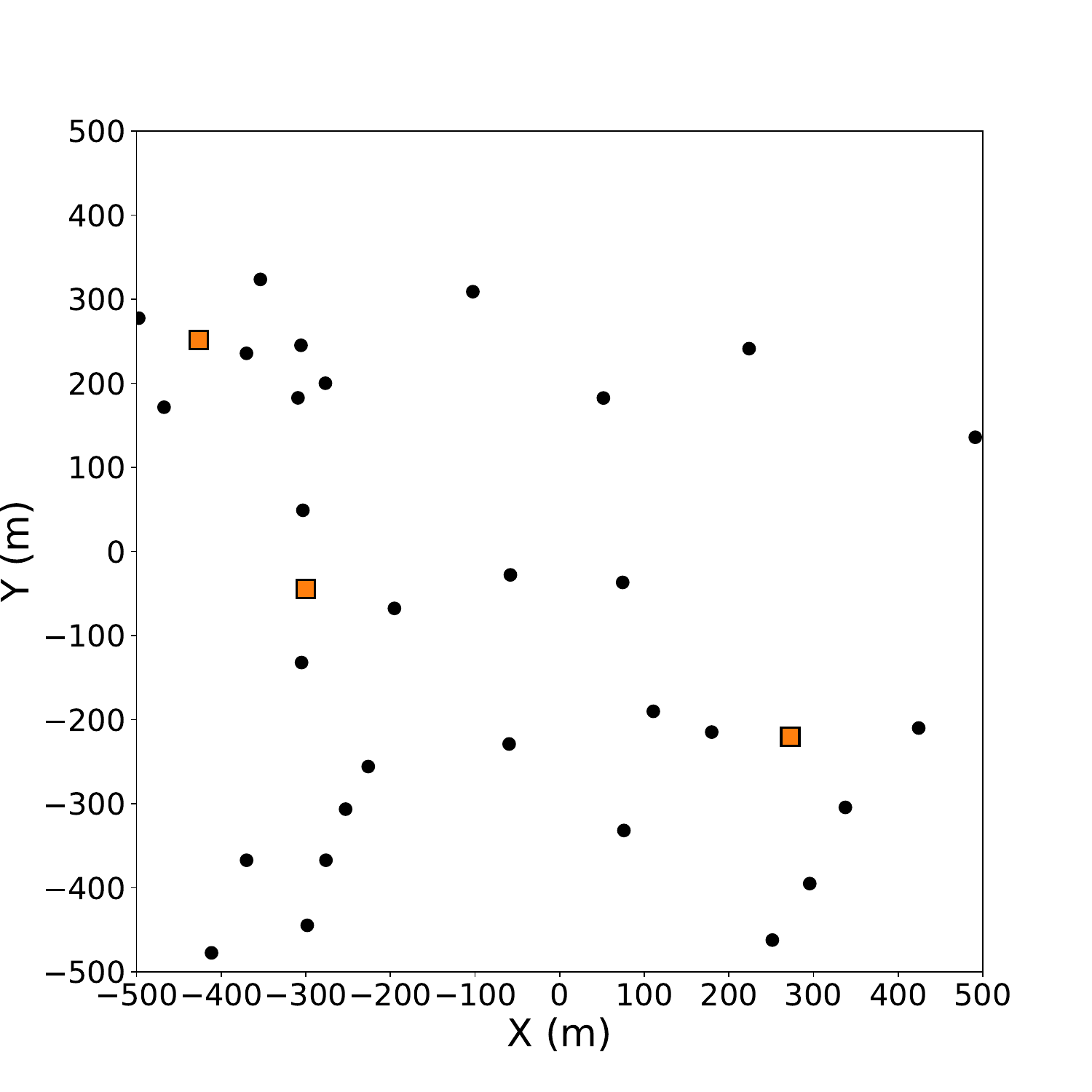}
    \subcaption{uniform}
  \end{minipage}
  \begin{minipage}[b]{0.45\linewidth}
    \centering
    \includegraphics[keepaspectratio, scale=0.3]{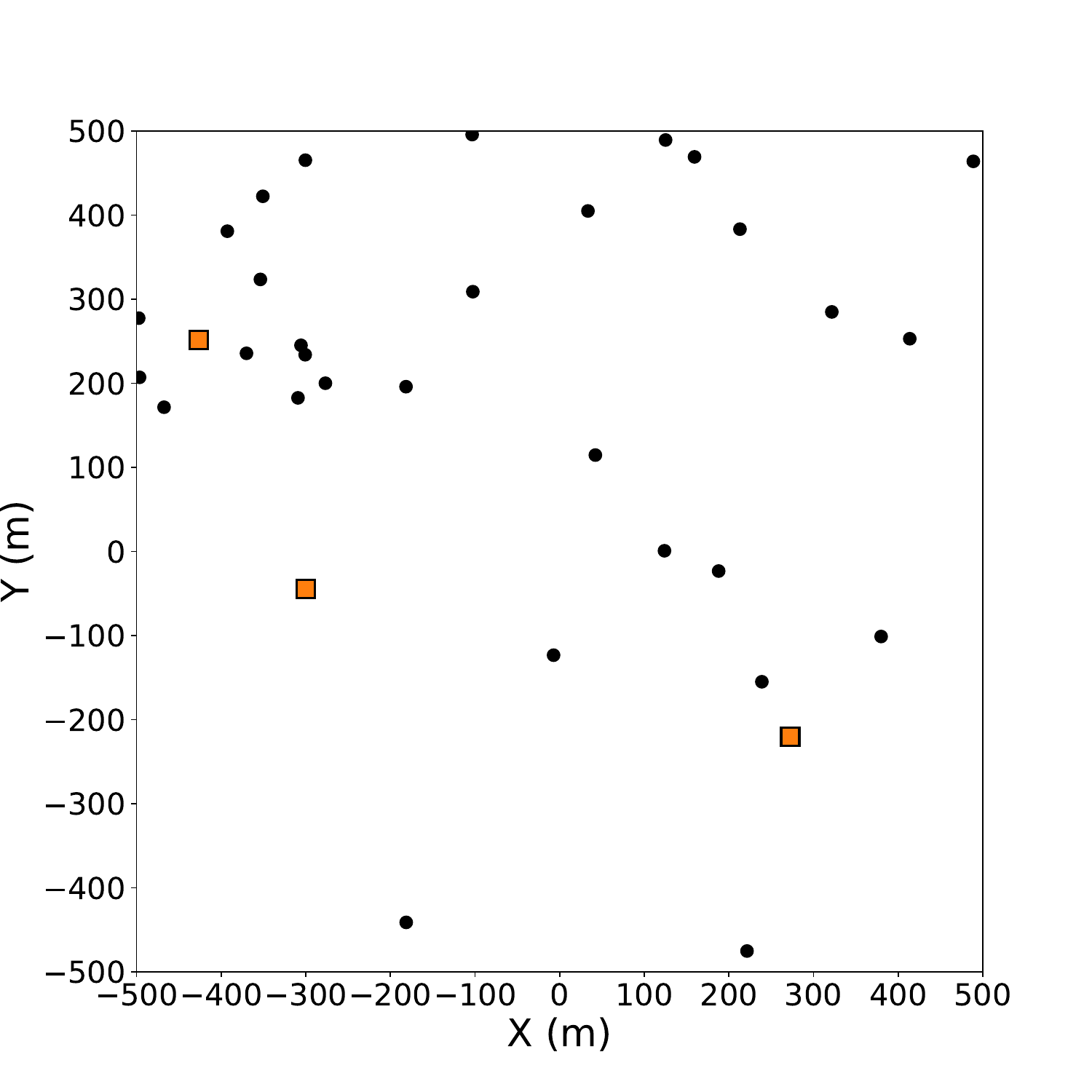}
    \subcaption{biased}
  \end{minipage}
  \caption{Examples of two mobile phone distribution patterns. The left figure shows a uniform distribution within the area, while the right figure shows a biased distribution towards a specific base station.}
  \label{example of distribution of mobile phones}
\end{figure}

\addcontentsline{toc}{section}{Experiments}
In this chapter, we show the results of the experiments to compare the naive formulation \eqref{naive qubo} with the proposed formulation \eqref{proposed qubo}. 
As experimental conditions, we randomly distribute mobile phones and base stations within an area. 
We conduct experiments under two scenarios: one where the distribution of mobile phones is uniform and another where the distribution is biased toward certain base stations. In a biased pattern, 60\% of mobile phones are located around one of the $M$ base stations, meaning that for 60\% of the mobile phones, the distance to this particular base station is the shortest. 
Examples of these distribution patterns are shown in Fig. \ref{example of distribution of mobile phones}.

In this experiment, we use the isotropic and Gaussian beam patterns shown in Fig. \ref{SINR heat map}. Therefore, we examine four test patterns derived from the combination of mobile phone placement and beam patterns, as shown in Table 1.

\begin{table}[htbp]
  \begin{center}
  \caption{Four test patterns in experiments.}
  \begin{tabular}{llllr}
    \hline
    test pattern	& distribution of mobile phones	& beam \\
    \hline
    1	& uniform	& isotropic \\
    2	& biased	& isotropic \\
    3	& uniform	& Gaussian \\
    4	& biased	& Gaussian \\
    \hline
  \end{tabular}
  \end{center}
\end{table}

First, we compare the number of qubits required by the two formulations for the same problem size of mobile phones and base stations. The hardware graph for embedding the QUBO is the Pegasus graph installed on the D-Wave Advantage 6.4. We consider two mapping techniques on the hardware graph. The first technique is minor embedding \cite{cai2014practicalheuristicfindinggraph}, which searches for chains of physical qubits that realize the logical couplings while heuristically minimizing the average chain length. The second technique is clique embedding, which searches for an embedding of the complete graph with the same number of vertices as the target graph.
 
\begin{figure}[htbp]
  \centering
    \includegraphics[width=0.9\linewidth]{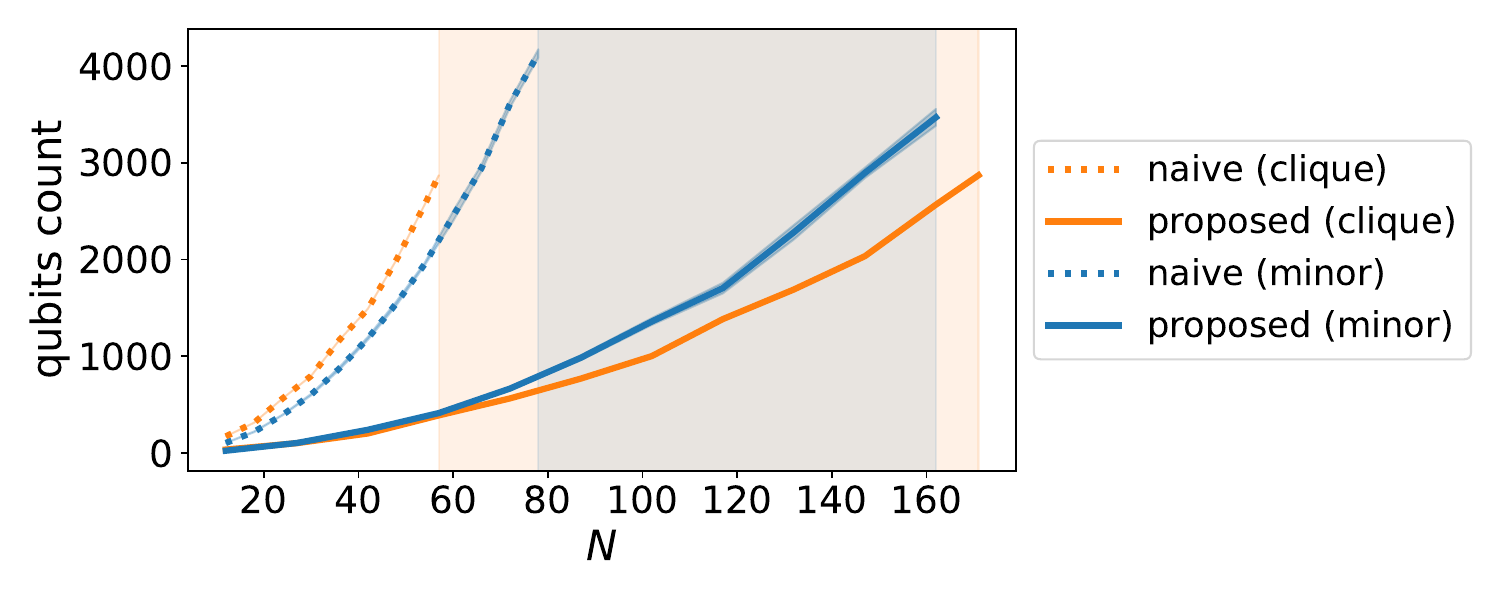}
    \caption{Dependency of the number of qubits required by the two formulations on the number of mobile phones when the number of base stations is fixed at 3. Each line plots the average of 10 instances. Orange lines correspond to clique embedding, blue lines to minor embedding.
    The points where the plot is cut off indicate the maximum number of mobile phones $N$ that could be embedded. The area with the colored background highlights the region of $N$ where only the proposed formulation can be embedded.}
\label{qubit comparison}
\end{figure}

We generate 10 random configurations of mobile phones and base stations and construct QUBOs for each, then compare the number of qubits required. In this experiment, we fix the number of base stations to 3 and examine how the number of required qubits increases with the number of mobile phones. The results are shown in Fig. \ref{qubit comparison}.

From Fig. \ref{qubit comparison}, it is clear that the proposed formulation requires fewer qubits than the naive formulation. Therefore, the proposed formulation is more efficient in using qubits. Moreover, in the proposed formulation, the clique-embedding curve stays below the hardware limit for a larger range of $N$ than the two minor-embedding curves. By contrast, in the naive formulation, the minor-embedding curve continues beyond the point where the clique-embedding curve reaches the device limit.
This pattern indicates that, for larger problem sizes, the naive formulation can remain embeddable through the flexibility of minor embedding.

Next, we compare the accuracy of the solutions obtained from the D-Wave quantum annealer using the two formulations. We use the D-Wave Advantage 6.4 for QA. As an embedding strategy, we employ the minor-embedding technique provided by D-Wave Systems’ minorminer library. After taking samples from the quantum processor, the solution is refined in a post-processing step to match the original QUBO problem. When chain breaking occurred, which refers to the case where qubits representing the same logical variable had different values, a majority voting strategy is applied to resolve the conflict. We obtained 1000 samples and selected the feasible solution with the lowest cost. The number of mobile phones is fixed at $N=30$, and the number of base stations at $M=3$. Each base-station transmits three identical Gaussian beams whose boresights are spaced at 120° intervals. In the experiments, we fix the boresight offsets of the three stations to 0°, 30°, and 60°, respectively.

\begin{figure}[htbp]
    \centering
    \includegraphics[width=0.9\linewidth]{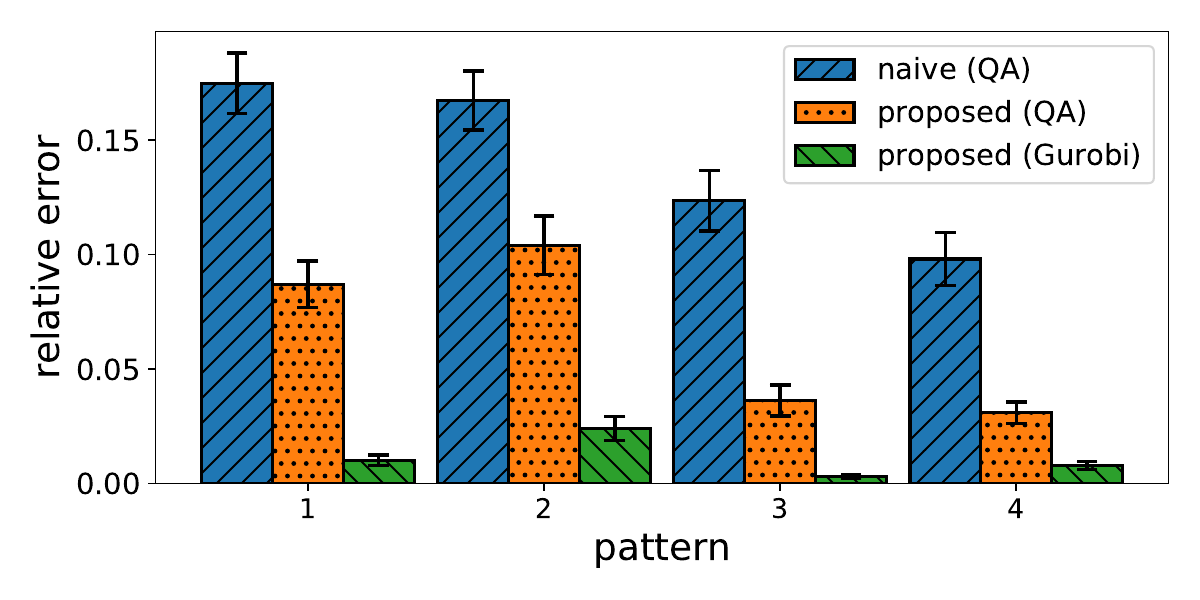}
    \caption{Comparison of the accuracy of solutions obtained using the two formulations across four test patterns. "naive (QA)" and "proposed (QA)" denote the solutions obtained by running the quantum annealer on the naive and proposed formulations, respectively. "proposed (Gurobi)" refers to the exact solution of the proposed formulation \eqref{proposed objective function} computed with Gurobi.}
    \label{4pattern}
\end{figure}

We conduct the experiment using the four test patterns shown in Table 1. We create 100 random configurations of mobile phones and base stations for each test pattern and solve each problem on the D-Wave quantum annealer. To distinguish the heuristic limitations of the quantum annealer from the approximation error introduced by our formulation \eqref{proposed objective function}, we also compute the exact optimal value and compare it with the QA results. For this purpose, we use the Gurobi Optimizer (version 10.0.0) as an exact solver. The results are shown in Fig. \ref{4pattern}.

The horizontal axis in Fig. \ref{4pattern} represents the four test patterns, and the vertical axis represents the relative error $(E - E^{\ast})/E^{\ast}$. Here, $E$ is the cost of the solution $x$ obtained from each method, given by $\sum_{i=1}^{N}  \left(S^{'}_{i,1} x_{i} + S^{'}_{i,2} (1 - x_{i}) \right)$, and $E^{\ast}$ is the cost of the exact solution $x^{\ast}$ of Eq. (8) obtained using a exact solver Gurobi, given by $\sum_{a=1}^{M} S_{i,a} x_{i,a}^{\ast}$. 
From Fig. \ref{4pattern}, it can be seen that the proposed formulation with quantum annealer results in smaller relative errors than the naive formulation in all test patterns. 
This indicates that the proposed formulation provides more accurate solutions when using a quantum annealer. Solving the proposed formulation \eqref{proposed objective function} with Gurobi yields solutions that are even closer to the true optimum. This observation confirms that an exact solution of the proposed formulation does not necessarily coincide with the global optimum of the naive formulation \eqref{naive objective function}, which enumerates all possible connection patterns. In addition, it can be observed that the accuracy is higher when the beam pattern is Gaussian (test patterns 3 and 4) compared to when it is isotropic (test patterns 1 and 2).

\begin{figure}[htbp]
    \centering
    \includegraphics[width=0.9\linewidth]{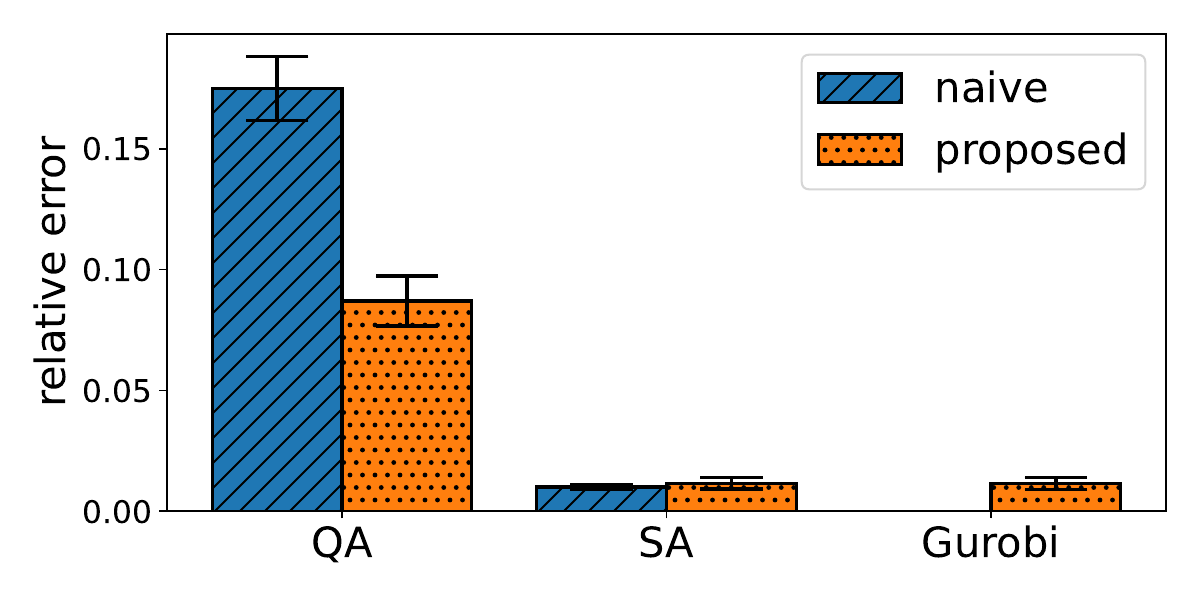}
    \caption{Comparison of the two formulations when solved using SA, QA, and Gurobi in test pattern 1. The bar labelled "naive" of Gurobi lies at zero relative error because this value serves as the reference $E^{\ast}$.}
    \label{sa_qa_comparison}
\end{figure}

Next, we compare the two formulations when solved using SA, QA, and Gurobi. For SA, we used the dwave.neal software (version 0.6.0) provided by D-Wave Systems \cite{dwave-neal}. The inverse-temperature schedule of SA was tuned on a subset of instances and fixed to $(\beta_{\rm start}, \beta_{\rm end})=(0.05, 5)$. 1000 samples were obtained, selecting the solution with the lowest cost. From the results in Fig. \ref{4pattern}, where the proposed formulation outperformed the naive formulation in all test patterns, we focus on test pattern 1 in the following experiments. 
The results are shown in Fig. \ref{sa_qa_comparison}.

The QA results in Fig. \ref{sa_qa_comparison} are identical to those in Fig. \ref{4pattern} for test pattern 1. 
Compared to QA, SA yields significantly lower relative errors for both formulations. 
This indicates that SA provides more accurate solutions than QA, which is susceptible to errors due to the physical device. 
On the other hand, unlike QA, the accuracy of the solutions obtained using the two formulations in SA is almost identical. Moreover, the accuracy obtained by applying SA to the proposed formulation \eqref{proposed qubo} is virtually identical to the Gurobi result, indicating that SA reaches solutions that are nearly optimal for this formulation.

\begin{figure}[htbp]
    \centering
    \includegraphics[width=0.9\linewidth]{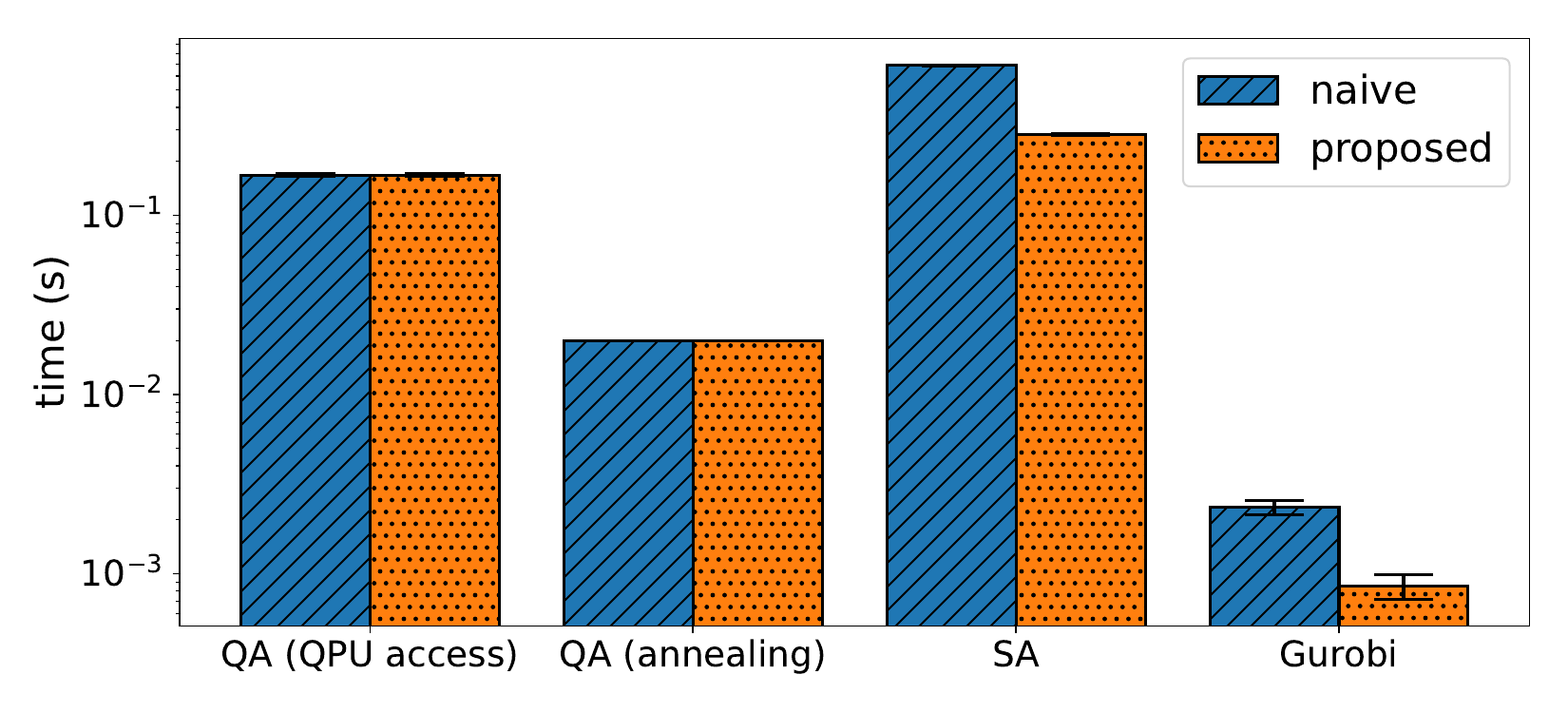}
    \caption{Comparison of computation times when solving the naive and proposed formulation with QA, SA, and Gurobi in test pattern 1 ($N=30,M=3$). "QA (QPU access)" denotes the QPU access time when performing QA, and "QA (annealing)" refers to the annealing time.}
    \label{time_comparison}
\end{figure}

Figure \ref{time_comparison} reports the computation times for test pattern 1 ($N=30, M=3$) when the naive and proposed formulations are solved with QA, SA, and Gurobi. For QA, we display both the annealing time and the quantum processing unit (QPU) access time. The annealing time equals the product of the per-sample annealing duration and the number of samples and is fixed to $20 \times 1000 = 0.02\ \mathrm{s}$ in this experiment. The QPU access time is the sum of the QPU sampling time and the QPU programming time. The sampling time aggregates, over all samples, the annealing time, the read-out time, and the hardware delay. The programming time occurs only once per problem instance and corresponds to configuring the QUBO on the chip. We use the annealing time as a measure of the theoretical performance of QA and the QPU access time as the latency that would be observed in a real application.

Gurobi returns solutions in the millisecond range and is therefore the fastest method, regardless of which formulation is used.
In contrast, SA requires between one and two orders of magnitude more time, making it the slowest of the three solvers.
QA occupies an intermediate position. The fixed "annealing time" is only $0.02\ \mathrm{s}$, yet the practical latency rises to roughly $0.2\ \mathrm{s}$ once QPU‐access overhead is included, so QA still trails well behind Gurobi while outperforming SA in wall-clock times.
A further pattern is that, for the classical algorithms (SA and Gurobi), the naive formulation consistently takes longer to solve than the proposed formulation, indicating that the variable-reduction scheme effectively shrinks the search space and shortens the solver’s internal exploration.
By contrast, the QA timings are almost identical for the two formulations, confirming that they are dominated by fixed hardware latencies rather than by the logical problem size in this range.

\begin{figure}[htbp]
    \centering
    \includegraphics[width=1.0\linewidth]{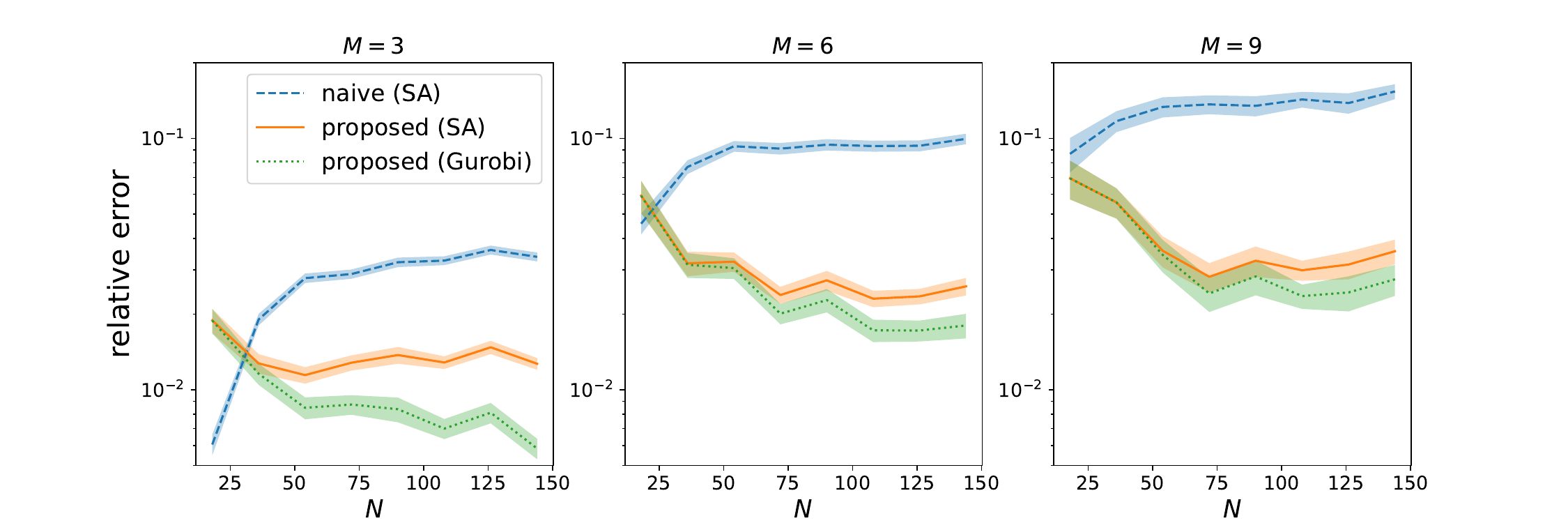}
    \caption{Dependence of the relative error of the two formulations in SA on the number of mobile phones. The three panels correspond to $M=3$ (left), $M=6$ (center), and $M=9$ (right).}
    \label{sa_n_dependence}
\end{figure}

We next investigate how the accuracy of solutions changes as the number of mobile phones and base stations increases in SA for both formulations. 
In this case, the number of base stations $M$ is fixed at 3, 6, and 9. The SA settings remain the same as in the previous experiment. The results are shown in Fig. \ref{sa_n_dependence}.

From the left panel of Fig. \ref{sa_n_dependence}, which corresponds to the case with $M=3$, it can be observed that the naive formulation \eqref{naive qubo} is slightly more accurate when $N$ is very small, but its error rises quickly once $N$ passes about 30. Beyond that point, the proposed formulation is more accurate. As the number of base stations increases with $M=6$ and $M=9$, the gap widens. The error of the naive formulation exceeds 10\ \%, whereas the proposed formulation remains below this threshold across the entire $N$ range.
In every setting, the exact result from Gurobi lies lower, which confirms that the remaining error comes from the heuristic search rather than from the formulation itself. These observations indicate that the proposed formulation maintains good accuracy as both the number of mobile phones and the number of base stations grow, while the naive formulation loses accuracy at larger scales.

\section*{Conclusion}
\addcontentsline{toc}{section}{Conclusion}
In this study, we conducted experiments comparing the naive formulation \eqref{naive qubo} with the proposed formulation \eqref{proposed qubo} in the context of optimizing connection patterns of mobile phones to base stations. In the experiment counting the number of qubits shown in Fig. \ref{qubit comparison}, we confirmed that when the number of base stations $M$ is 3, the number of mobile phones that can be handled on the D-Wave Advantage increases. This is because the number of logical variables required by the proposed formulation is reduced to $1/M$ of that required by the naive formulation. By reducing the number of logical variables, the proposed formulation can handle problems with larger mobile phones. These results indicate that the proposed formulation effectively reduces the required qubits. 

The same figure also reveals a qualitative difference between clique embedding and minor embedding. For the proposed formulation, the curve obtained with clique embedding remains feasible up to much larger $N$ than the curve obtained with minor embedding, whereas for the naive formulation, the opposite occurs, and only minor embedding remains feasible at larger $N$. This is because clique embedding searches for an embedding of a complete graph whose number of nodes equals the number of QUBO variables, making the embedding cost independent of the QUBO’s density. By contrast, minor embedding seeks an embedding tailored to the specific connectivity graph of the QUBO, so the required resources depend on how dense that graph is.

In the subsequent experiment, as shown in Fig. \ref{4pattern}, we compared the accuracy of solutions obtained from the D-Wave quantum annealer across the four patterns listed in Table 1. The results confirmed that the proposed formulation consistently outperforms the naive formulation regarding solution accuracy across all four patterns. The reason for this improvement is likely because, as mentioned above, the proposed formulation reduces the number of logical variables to $1/M$ of that in the naive formulation, reducing the number of qubits required. Since the D-Wave quantum annealer is susceptible to external noise, the more qubits are controlled during QA execution, the greater the likelihood of errors. Therefore, reducing the number of qubits in the proposed formulation improves solution accuracy. Thus, the proposed formulation effectively uses limited computational resources on the D-Wave quantum annealer, providing good approximate solutions. In addition, solving the proposed formulation \eqref{proposed objective function} with Gurobi gave an even lower error than QA, which represents the optimal value of this model. This result demonstrates that an exact solution of the proposed formulation does not always coincide with the global optimum of the full problem \eqref{naive objective function} because the reduced model ignores all but the first and second candidate base stations for each mobile phone.

Figure \ref{time_comparison} shows a comparison of computation time. QA showed an intrinsic annealing time of $0.02\ {\rm s}$ and a practical latency of about $0.2\ {\rm s}$ once QPU access overhead was included. Although this latency is larger than that of Gurobi, it remains well below the re-optimization window required in realistic deployments. In this paper, we considered a scenario where the positions of the mobile phones are fixed within the area. However, in real-world scenarios, mobile phone positions change over time as people move. 
Therefore, the connection pattern optimization needs to be repeated periodically. Assuming a cell diameter of about $500\ {\rm m}$ and a typical vehicle speed of $60\ {\rm km/h}$, a mobile phone crosses a cell boundary in roughly $30\ {\rm s}$, so optimization must complete within that interval. The observed QPU access time satisfies this requirement, and because the annealing time is essentially independent of the logical problem size, it is expected to remain within the window even for larger optimization areas.

Moreover, as shown in Fig. \ref{sa_n_dependence}, the proposed formulation also outperforms the naive formulation regarding solution accuracy as the number of mobile phones increases, even when using the classical algorithm SA. When the number of base stations increases to $M=6$ and $M=9$, the gap becomes larger. The proposed formulation also showed a gradual increase in error, even when solved exactly with Gurobi, because it still considers only the two best base stations per mobile phone, so the approximation becomes coarser when each mobile can see more than two good candidates.

However, as the number of mobile phones and base stations increases, the number of possible solutions increases exponentially, making naive formulation a challenging combinatorial optimization problem. Therefore, solving the QUBO based on this becomes increasingly difficult with the increase in problem sizes, even in SA, where there is no noise influence as in QA. On the other hand, the proposed QUBO is a simplified problem in which each mobile phone is only connected to the best or second-best base station, enabling stable solutions even as the problem size increases.

In conclusion, the proposed formulation effectively reduces the number of qubits and improves the accuracy of approximate solutions, making it an efficient approach for utilizing the limited computational resources of quantum annealers. 
In addition, even in classical QUBO solvers like SA, the proposed formulation becomes increasingly effective in providing good approximate solutions as the problem size grows.

Future work includes the following considerations. 
First, the proposed formulation was originally introduced in the context of evacuation optimization. 
In this study, we applied it to optimize connection patterns to base stations. 
The key idea of the proposed formulation is to simplify the problem by allowing choices between the best and second-best options, which we believe has potential for various other applications in real-world scenarios. 
We plan to explore other possible applications of this formulation.

In addition, while the proposed formulation can reduce the number of qubits required and handle more mobile phones compared to the naive formulation, the number of mobile phones is much larger in real situations. 
In current systems, a single macro base station usually serves three fixed antenna sectors. When one sector is chosen as the center of an optimization area, the first six closer neighbors form a cluster of seven sectors, as in the literature \cite{s21196608}. Assuming that each sector supports roughly $300$ simultaneously active mobile phones, which yields about 2100 mobile phones in one optimization area. Even if each mobile is allowed to choose only its best and second-best base station, the corresponding QUBO would contain several thousand logical variables. The embedding results in Fig. \ref{qubit comparison} suggest that current Pegasus hardware can not map such a problem.
For this demand, two countermeasures can be considered. One remedy is to partition the network into smaller sub-areas that contain fewer base stations and mobile phones. The proposed formulation is especially suitable for this approach because its accuracy remains stable when each sub-problem involves few base stations. Each sub-problem can then be solved independently on the quantum annealer, and the local solutions can be merged. 
A second remedy is to run a quick classical pre-processing step and forward only the more challenging subset to the quantum annealer. Using these two steps keeps the QUBO problem within the hardware limit and enables the quantum annealer to address the part of the problem in which it is expected to be valuable.
\bibliography{Reference}

\section*{Acknowledgments}
\addcontentsline{toc}{section}{Acknowledgments}
This study was financially supported by programs for bridging the gap between R\&D and IDeal society (Society 5.0) and Generating Economic and social value (BRIDGE) and Cross-ministerial Strategic Innovation Promotion Program (SIP) from the Cabinet Office.

\section*{Author contributions statement}
\addcontentsline{toc}{section}{Author contributions statement}
Takabayashi, T., performed the experiments and analyzed the results. Sudo, S., proposed the research topic and problem formulation and created a calculation tool for problem formulation. Aoki, T., and Seo, S., contributed to refining the research direction, provided input on interpreting the calculation results, and contributed to preparing the manuscript. Ohzeki, M., conceived the idea of the proposed method and supervised this work. 
All authors participated in discussions of the results and contributed to the final manuscript.

\section*{Additional Information}
\subsection*{Data Availability}
The datasets used during the current study are available from the corresponding author upon reasonable request.

\end{document}